\documentclass[pdflatex,%referee,
sn-vancouver-num]{sn-jnl}% Vancouver Numbered Reference Style

\usepackage{graphicx}%
\usepackage{multirow}%
\usepackage{amsmath,amssymb,amsfonts}%
\usepackage{amsthm}%
\usepackage{mathrsfs}%
\usepackage[title]{appendix}%
\usepackage{xcolor}%
\usepackage{textcomp}%
\usepackage{manyfoot}%
\usepackage{booktabs}%
\usepackage{algorithm}%
\usepackage{algorithmicx}%
\usepackage{algpseudocode}%
\usepackage{listings}%
\usepackage{cleveref,physics,siunitx}
\DeclareSIUnit\Gauss{G}
\begin{document}

\title[Cold-Atom Buoy]{Cold-Atom Buoy: A Differential Magnetic Sensing Technique in Cold Quadrupole Traps} % brewed together with Thomas

\author[1]{\fnm{Árpád} \sur{Kurkó}}

\author[1]{\fnm{Dávid} \sur{Nagy}}

\author[1,4]{\fnm{Alexandra} \sur{Simon}}

\author[1]{\fnm{Thomas W.} \sur{Clark}}

\author[1]{\fnm{András} \sur{Dombi}}

\author[1,4]{\fnm{Dániel} \sur{Varga}}

\author[1]{\fnm{Francis B.} \sur{Williams}}

\author[2]{\fnm{József} \sur{Fortágh}}

\author[1,3]{\fnm{Peter} \sur{Domokos}}

\author*[1]{\fnm{András} \sur{Vukics}}\email{vukics.andras@wigner.hun-ren.hu}

\affil[1]{\orgname{HUN-REN Wigner Research Centre for Physics}, \orgaddress{\street{Konkoly-Thege M. út 29-33}, \city{Budapest}, \postcode{1121}, \country{Hungary}}}

\affil[2]{\orgdiv{Center for Quantum Science, Physikalisches Institut}, \orgname{Eberhard Karls Universität Tübingen}, \orgaddress{\street{Auf der Morgenstelle 14}, \city{Tübingen}, \postcode{72076}, \country{Germany}}}

\affil[3]{Department of Theoretical Physics, Institute of Physics, Budapest University of Technology and Economics, H-1111 Budapest, Műegyetem rkp. 3, Hungary}

\affil[4]{Department of Physics of Complex Systems, ELTE Eötvös Loránd University, Pázmány Péter sétány 1/A, H-1117 Budapest, Hungary}


\abstract{We present a differential technique for vector magnetic sensing based on a cold-atom cloud in a magnetic quadrupole trap. An external homogeneous magnetic field displaces the trap center in a direction and magnitude proportional to the field. By reversing the quadrupole polarity between experimental shots and comparing the resulting cloud positions, we extract a differential displacement signal that is free from common-mode effects such as gravity and weak magnetic-field inhomogeneities. The signal is directionally proportional to the external field and requires only absorption imaging, without spectroscopic interrogation. Assuming micron-scale position resolution, the technique enables field resolution at the milli-Gauss level. It offers a practical tool for field compensation in magnetically sensitive experimental stages, bridging operational regimes from Earth-level fields to atomic magnetometry. A straightforward extension to full three-dimensional sensing is possible with only a minimal addition to standard cold-atom infrastructure.}




\maketitle

\section{Introduction}\label{sec1}
Magnetic trapping is a foundational technique in cold-atom physics, enabling conservative, long-duration confinement of neutral atoms in magnetically sensitive internal states \cite{Metcalf1999MagneticTrapping}. Following the advent of magneto-optical traps (MOTs) \cite{Raab1987}, the development of purely magnetic traps allowed atomic ensembles to be held and manipulated without continuous optical scattering. This led to the first observations of Bose–Einstein condensation \cite{Anderson1995,Davis1995} and has since underpinned advances in quantum simulation, interferometry, and hybrid quantum systems \cite{Bloch2008ManyBodyPhysics,Jo2007LongPhaseCoherence,Verdu2009StrongMagneticCoupling}. The canonical magnetic trap geometries – such as the quadrupole and the Ioffe–Pritchard configuration – are typically realized using either macroscopic coils mounted near or inside the vacuum chamber, or micrometer-scale current paths on atom chips \cite{Petrich1995,Reichel1999,Fortagh2007}. These setups exploit the spatial dependence of the magnetic field near the trap center to create a potential minimum for low-field-seeking Zeeman states. A central advantage of such magnetic traps is that they are easy to configure (e.g. displace) via the feed current of the electromagnets \cite{gehm1998dynamics,greiner2001magnetic,Folman2002MicroscopicMagneticTraps,liu2023ultra}.

Atomic systems account for some of the most sensitive \cite{kominis2003subfemtotesla,dang2010ultrahigh,sheng2013subfemtotesla} and accurate \cite{farooq2020absolute} means of magnetic field measurement, having become competitive with superconducting quantum interference devices \cite{buchner2018tutorial}. These sensors rely on the internal degrees of freedom of atoms – typically hyperfine Zeeman sublevels – and detect magnetic fields through their influence on atomic spin precession via the optical response \cite{BudkerKimball2013}. A wide variety of architectures exist, ranging from thermal atomic beams used in early precision experiments \cite{rabi1938new}, through spin-exchange relaxation-free magnetometers based on alkali vapors \cite{bell1961optically, Kominis2003, BudkerRomalis2007, Allred2002, Shah2007}, to laser-cooled atomic ensembles in vacuo \cite{Vengalattore2007, Afach2015}. Recent developments include all-optical multi-axis measurements \cite{li2024all} as well as multi-dimensional magnetic sensing using spatially shaped beams \cite{castellucci2021atomic}.

In this work, we introduce a technique for magnetic field sensing based on the spatial displacement of a cold-atom cloud held in a magnetic quadrupole trap and manipulated exclusively with changes in current. By alternating the polarity of the quadrupole field between consecutive experimental shots, we generate an antisymmetric displacement response of the trap center in the presence of an external homogeneous magnetic field. This effect is measured using absorption imaging and two-dimensional Gaussian fitting of the atom cloud’s optical depth profile.

The technique is remarkably simple: it relies solely on spatial degrees of freedom and requires no spectroscopic interrogation, microwave fields, or internal-state coherence – making it broadly applicable in settings where magnetic field control is critical but conventional magnetometry is impractical, as for magnetic compensation in cold-atom preparation stages. The buoy technique uses the same cold-atom sample and the same trapping configuration as those employed in the target experiment, the magnetic-field information being obtained exactly at the position of the atoms for which field compensation or characterization is required. Relying on just a single imaging beam, the technique can facilitate portable \cite{bidel2013compact,ehinger2022comparison} and miniature \cite{nshii2013surface,nichols2020magneto,mcgilligan2020laser} devices in cold-atom-based quantum technologies.

The strategy of reversing a controlled experimental parameter to reveal or isolate physical effects is a cornerstone of precision metrology. It appears across a wide range of domains: in spin-echo and Ramsey spectroscopy sequences that cancel quasi-static dephasing \cite{Hahn1950SpinEcho,Ramsey1950SeparatedOscillatoryFields}; in electric dipole moment searches where field polarity is flipped to isolate parity-violating signatures \cite{roussy2023improved}; and in Pound-Drever-Hall spectroscopy \cite{drever1983laser} where opposite-sign phase shifts of sidebands cancel in the midpoint of a spectral feature. These differential schemes amplify antisymmetric responses while canceling common-mode drifts and offsets, enabling high-precision measurements in the presence of substantial technical noise. The cold-atom buoy technique follows this paradigm by alternating the sign of the quadrupole gradient between consecutive experimental shots 
%– not to manipulate internal states, because according to the adiabaticity, the atoms always remain in the same, low-field seeking state – but
to reverse the geometric response of the trap to external homogeneous magnetic fields. The resulting displacement difference forms a direct observable that depends linearly and directionally on the external field.

\section{Concept}

\subsection{Scheme}
The potential in a magnetic trap acting on an atom in a magnetic sublevel $m_F$ reads:
\begin{equation}
\label{eq:potential}
    U(\vb{r}) = \mu_\text{B} g_F m_F \abs{\vb{B}(\vb{r})},
\end{equation}
where $\mu_\text{B}$ is the Bohr magneton, and $g_F$ is the hyperfine Landé g-factor. The potential being proportional to the modulus of the magnetic field shows that the trapping is independent of the polarity of the quadrupole, in contrast to a MOT, where the quadrupole polarity has to match the polarization of the inbound MOT beams.

The response to a homogeneous field, on the other hand, does depend on the quadrupole polarity, and is opposite for opposite quadrupole polarities, forming the basis of the buoy effect. This can be seen by writing the quadrupole field in the following form:
\begin{equation}
\label{eq:Qpole}
\vb{B}(\vb r) = \vb Q\,\vb r,\qqtext{with}\vb Q\equiv Q\begin{pmatrix}1 & 0 &0\\0 &1 &0\\0 &0 &-2\end{pmatrix},\qqtext{and}
Q \equiv \hat{\vb x} \cdot \partial_x \vb B(\vb r)\big\vert_{\vb r=\epsilon\hat{\vb x}}
\end{equation}
where $Q$ is the strength of the quadrupole, considered a signed scalar quantity, that can be expressed via the gradient of the field in one of the Cartesian directions at an infinitesimal distance $\epsilon$ from the center. The sign of $Q$ corresponds to the quadrupole polarity: positive when the axial field gradient points inward along $+z$, and negative otherwise.

While this form of the field leads to a potential with its minimum at $\vb 0$, where $\vb{B}=0$, the superposition of a homogeneous external field $\vb{B}_\text{ext}$ displaces this zero-field point to
\begin{subequations}
\begin{equation}
\label{eq:displacedCenter}
    \vb r_0 = -\vb Q^{-1}\,\vb{B}_\text{ext}+\vb 0, \qqtext{with}
    \vb Q^{-1}=\frac1{2Q}\begin{pmatrix}2 & 0 &0\\0 &2 &0\\0 &0 &-1\end{pmatrix}
%    = \frac{\qty[ \hat{\vb z}\otimes \hat{\vb z} - 2 \qty(\hat{\vb x}\otimes \hat{\vb x} + \hat{\vb y}\otimes \hat{\vb y} ) ]\,\vb{B}_\text{ext}}{2 Q} + \vb 0.
\end{equation}
The somewhat unusual-looking addition of the origin vector – the position of the quadrupole center with no external field – is intended to emphasize that this is an unknown: the origin $\vb 0$ can be different from a geometrical center in the experimental setup due to imperfections of the assembly (e.g. coil windings), and it cannot be directly observed, since we can never fully isolate the system from external fields. Therefore, $\vb r_0$ cannot be measured directly, however, notice that the first term in its expression changes sign with $Q$, whereas the origin remains unaffected, i.e. any uncertainty in the position of the origin is a common mode under quadrupole polarity reversal.

\begin{figure}
    \centering
    \includegraphics[width=\linewidth]{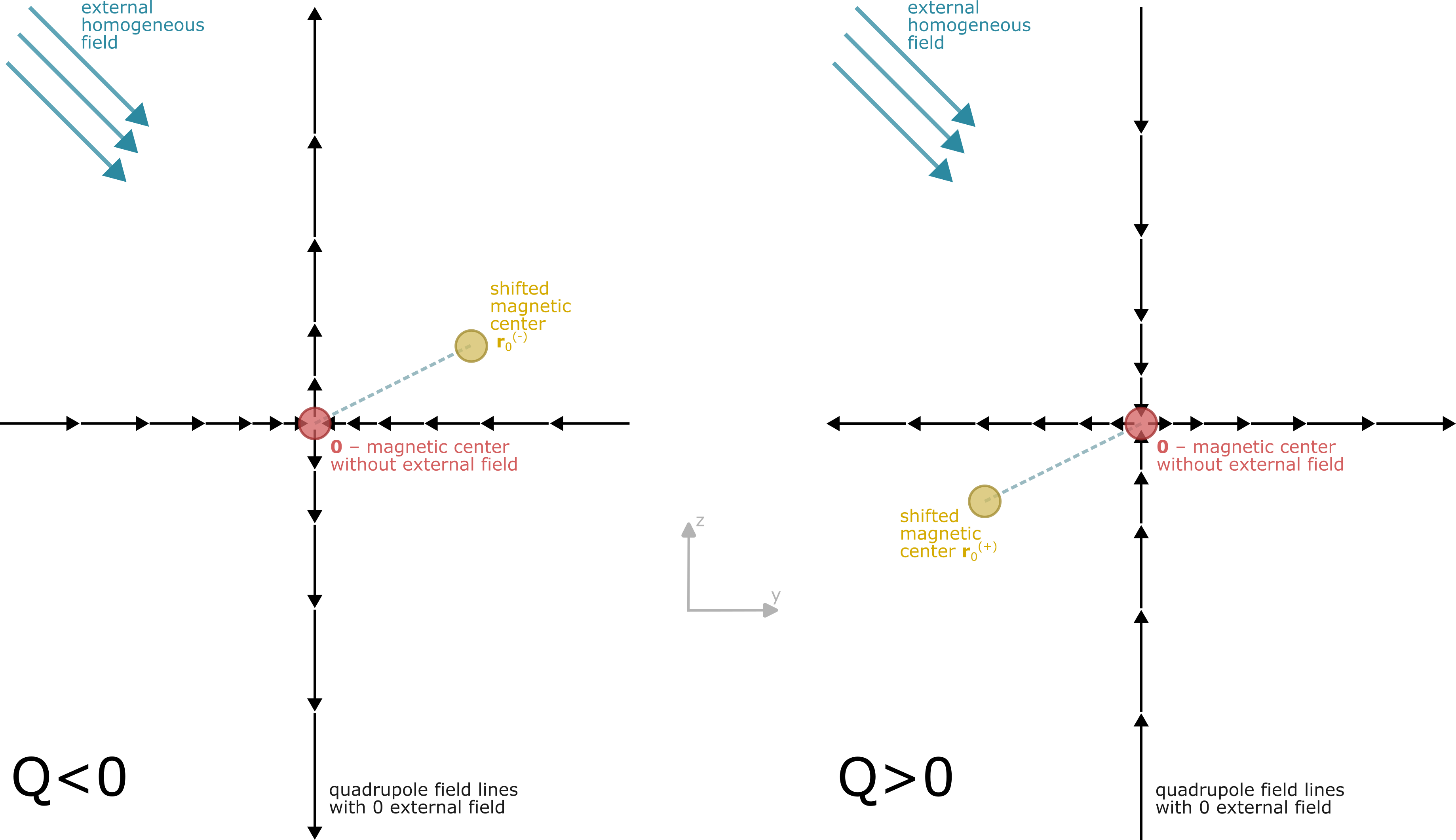}
    \caption{\textbf{Schematic of the cold-atom buoy concept.} In the quadrupole configuration, the magnetic field grows linearly in the three Cartesian directions with the distance from the center (with opposite sign and doubled gradient in the vertical direction). This makes that an external homogeneous field shifts the center of the quadrupole (the $\vb B=0$ point – denoted with orange) with the same distance, but opposite directions from the origin (denoted with red circle) under a switch of quadrupole polarity. For the definition of $Q$, cf. \cref{eq:Qpole}. The buoy analogy is based on the identification of the shifting cold-atom cloud with the buoy floating on the surface of the water (the plane transverse to the imaging axis), and the zero-external-field magnetic center, $\vb 0$, with the anchor point of the buoy. The external field is the current that displaces the buoy.}
    \label{fig:buoyScheme}
\end{figure}

This observation underpins the cold-atom buoy technique, cf. \cref{fig:buoyScheme}, and motivates us to introduce
\begin{equation}
\vb r_0^{(+)} \equiv \vb r_0(Q>0), \qqtext{and} \vb r_0^{(-)} \equiv \vb r_0(Q<0),
\end{equation}
\end{subequations}
corresponding to the shifted trap centers for opposite quadrupole polarities. The relevant observable is the differential displacement
\begin{equation}
\label{eq:buoySignal}
\Delta \vb r \equiv \vb r_0^{(+)} - \vb r_0^{(-)}=-2\,\vb Q^{-1}\,\vb{B}_\text{ext},
\end{equation}
that depends linearly and directionally on $\vb B_\text{ext}$, and is free from the uncertainties of $\vb 0$, as well as other common-mode effects, including – as analyzed subsequently – gravity and weak external inhomogeneities. Conversely, the midpoint between the two displacements $\frac{\vb r_0^{(+)}+\vb r_0^{(-)}}2\approx\vb 0$ can be used as an estimate of the origin. This can be useful for MOT adjustments, however, as we show explicitly in \cref{fig:results1_small} this estimate is affected by common-mode imperfections.

Two remarks clarify the relation of the buoy technique to displacement-based magnetometry in MOTs. (1) The cold-atom buoy is a purely geometric effect that relies on the linear, directional dependence of the field \labelcref{eq:Qpole} in a quadrupole trap, in contrast to a magneto-optical trap, where the interplay between magnetism and light forces sets a much more complex scenario. (2) Moreover, MOT-center displacement measurements are inherently non-differential: the quadrupole polarity cannot be reversed independently of the optical configuration without destroying the trapping conditions. In addition, MOTs are generally less cold than magnetically trapped ensembles, leading to lower spatial resolution in the determination of the trap center.

\Cref{eq:Qpole} is an excellent approximation of the magnetic field of a coil pair with opposite driving currents in the vicinity of the geometrical midpoint between the coils,  cf. \cref{sec:precision}. Assuming perfectly balanced driving of the two coils making up the quadrupole, $Q$ is proportional to the driving current, so its sign can be flipped simply by flipping the direction of the drive current. The magnetic zero, being the minimum of the potential, is designated by the point of maximum density of the trapped atomic cloud, that can be determined by absorption imaging.

\subsection{Demonstration}
\label{sec:demonstration}
The following experiments are performed in a rubidium-87 cold-atom system, described in \cref{sec:system}. To experimentally characterize the buoy effect, the external homogeneous field in all three spatial directions is varied systematically. For this, we vary the currents in three coil pairs operated in the dipole configuration leading to closely homogeneous field in the region of interest. The external homogeneous field can therefore be treated as
\begin{equation}
\label{eq:Bext}
\vb{B}_\text{ext}=\vb{B}_\text{stray}+\vb{B}_\text{bias}(I_x,I_y,I_z),
\end{equation}
where $\vb{B}_\text{stray}$ encompasses fields not controlled by us, stemming from the Earth and nearby devices.

The shifted quadrupole center is determined by absorption imaging \cite{Andrews1997InterferenceBECs,Hueck2017HighResolutionImaging} of the magnetically trapped cloud, for details cf. \cref{sec:AI}. For every setting of the currents $(I_x,I_y,I_z)$, we record absorption images for both positive and negative quadrupole polarities, and extract $\vb r_0^{(+)}$ and $\vb r_0^{(-)}$.

In its pure form, the buoy effect enables sensing in the two directions transverse to the imaging axis which here is denoted by $x$. Moreover, according to \cref{eq:displacedCenter}, the displacement in each Cartesian direction is solely determined by the corresponding component of the external field; e.g., displacement along $x$ reflects only $B_{\text{ext},x}$, and so forth. Therefore, given a single imaging axis $x$, we gain information about the $y$ and $z$ components by varying the currents corresponding to those two directions.

\begin{figure}
    \centering
    \includegraphics[width=0.9\linewidth]{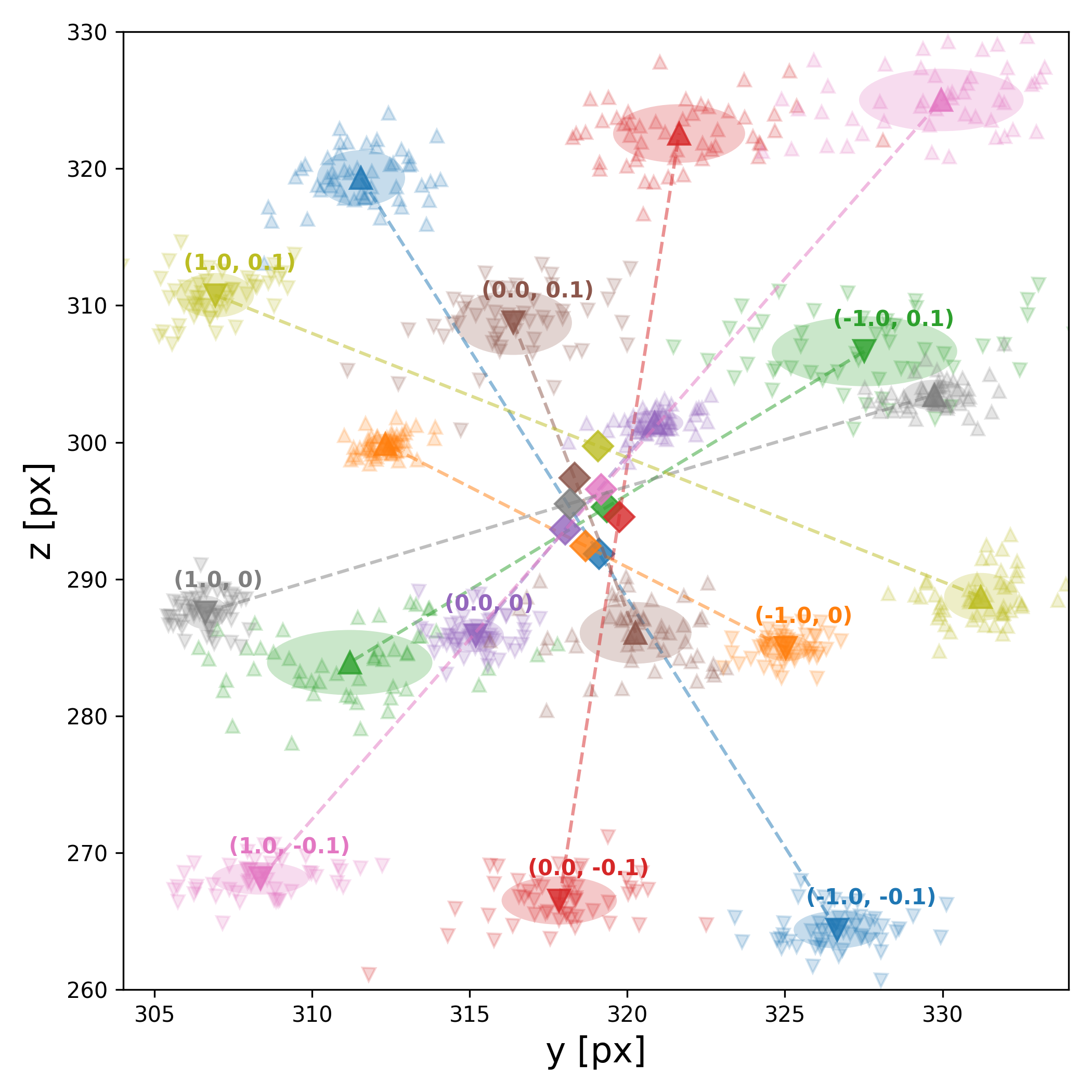}
    \caption{\textbf{Demonstration of the buoy effect.} Positions of the magnetically trapped cloud for various settings of the applied bias currents \( (I_y, I_z) \), displayed in color code on the \((y, z) \) image plane in the field of view of absorption imaging, measured in CCD pixels. For each \( (I_y, I_z) \) pair and each quadrupole polarity, 50 absorption images were recorded. Small triangular markers indicate the center-of-mass position of the cloud extracted from a two-dimensional Gaussian fit of each individual absorption image. Upward- and downward-pointing triangles denote shots with opposite quadrupole polarities. The identically colored label \((I_y, I_z)\) at each cloud cluster indicates the applied current pair in ampere for that dataset. For each \( (I_y, I_z) \) pair, the cluster centers corresponding to the two quadrupole polarities – denoted by identical, but larger triangles, and corresponding to estimates of $\vb r_0^{(+)}$ and $\vb r_0^{(-)}$ – are connected by a dashed line segment. The midpoint of this segment (shown as a large rhombus of the same color) is the inferred origin $\vb 0$ – the magnetic center of the quadrupole with zero external field. Identically colored ellipses around the large triangles represent standard deviations of the ensembles for each configuration. The difference between the scale of the currents \( I_y \) vs. \( I_z \) stems from the different geometry of the corresponding coil pairs, cf. \cref{sec:system}.}
    \label{fig:results1}
\end{figure}

\Cref{fig:results1} presents data obtained across a broad range of applied compensation currents $(I_y,I_z)$. Here, we performed 50 experimental shots per $(I_y,I_z)$ pair and quadrupole polarity. The figure shows the absorption imaging field of view. Each small triangle represents the center-of-mass position of the magnetically trapped cloud in a single shot, extracted via a 2D Gaussian fit to the optical density (OD) map, cf. \cref{sec:AI}. The color indicates the particular $(I_y,I_z)$ setting, while the triangle orientation (upward or downward) denotes the quadrupole polarity. Large triangles show the shot-averaged positions for each condition, i.e., the estimates of $\vb r_0^{(+)}$ and $\vb r_0^{(-)}$. The results clearly exhibit the buoy effect:
\begin{enumerate}
    \item The midpoint between the $\vb r_0^{(+)}$ and $\vb r_0^{(-)}$ estimates, i.e., the estimate of the origin $\vb 0$ (marked by a large rhombus) remains nearly invariant across different compensation current settings, as expected. The tight clustering of rhombuses confirms that the differential technique isolates the true magnetic center, independent of the externally applied fields.
    \item Varying \( I_y \) causes horizontal displacement, while varying \( I_z \) results in vertical displacement, confirming the expected directional response, although some residual displacement is observed in the orthogonal direction.
\end{enumerate}

In principle, coupling between directions could be attributed to imperfections of the alignment of the imaging plane and axis, or – as shown in \cref{sec:inhomogeneities} – to inhomogeneities of the stray field. The first leads to a differential signal, whereas the second is a common-mode effect to first order in the strength of the inhomogeneity cf. \cref{sec:CMR}. As we will demonstrate in \cref{fig:results1_small}, the directional coupling in our data is a common-mode effect, and hence can be attributed to the inhomogeneity, most plausibly that of the ion pump’s stray field.

\begin{figure}
    \centering
    \includegraphics[width=0.9\linewidth]{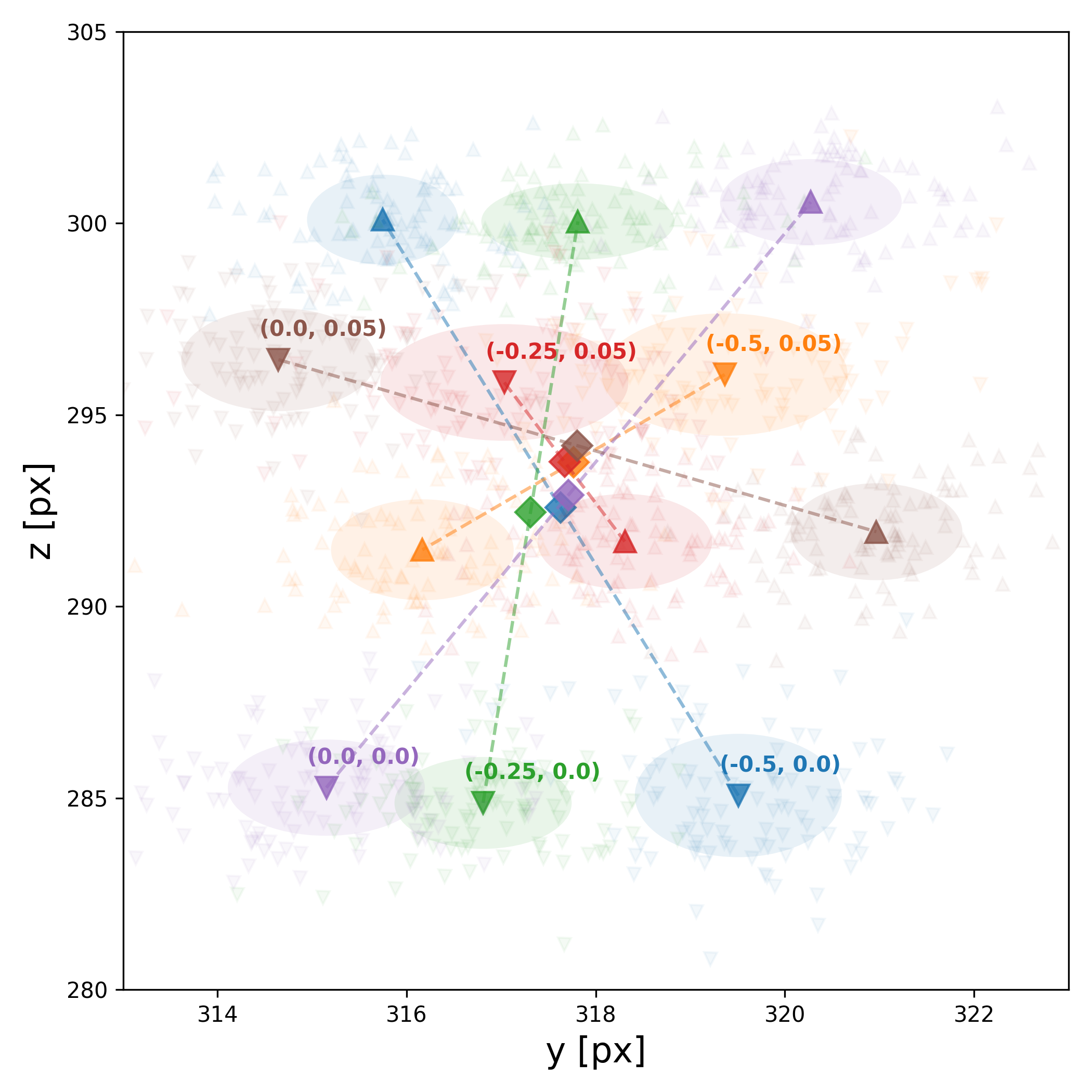}
    \caption{\textbf{Zoomed-in displacement data around the compensation point.} Same measurement as in \cref{fig:results1}, but recorded over a finer grid of $I_y$ and $I_z$ values centered around the compensation currents inferred from that figure. Here, 100 shots were taken for each current pair and quadrupole polarity. This finer sampling allows for a more precise localization of the origin $\vb 0$, contributes better data for determining the exact compensation values, and provides further validation of the buoy effect at higher resolution.}
    \label{fig:results2}
\end{figure}

\Cref{fig:results2} presents results across a finer range of $I_y$ and $I_z$ values centered around the compensation currents that can be inferred from \cref{fig:results1}. Here, 100 experimental shots were taken for each condition. This finer sampling enables a more precise identification of $\vb 0$, and contributes better data for determining the exact compensation values. 

Next, we consider $\Delta \vb r$ to exploit its directional \emph{and linear} response to the total external field $\vb B_\text{ext}$, as given by \labelcref{eq:buoySignal}. We also invoke the linear dependence of the controlled bias field $\vb B_\text{bias}$ on the applied coil currents. Hence the compensation currents that zero the net external field in each direction can be inferred via linear regression. That is, according to the Biot–Savart law, we can write
\begin{equation}
\label{eq:compensation}
\vb{B}_\text{bias} = \hat{\vb x}\,\alpha_x I_x + \hat{\vb y}\,\alpha_y I_y + \hat{\vb z}\,\alpha_z I_z,
\end{equation}
where the coefficients \( \alpha_i \) characterize the effective field contributions of the corresponding coil pair at the MOT center. Compensation means finding the currents where
\begin{equation}
\vb{B}_\text{bias}(I_x^@,I_y^@,I_z^@)=-\vb{B}_\text{stray}\qqtext{i.e.}
I_x^@=-\frac{\hat{\vb x}\cdot\vb{B}_\text{stray}}{\alpha_x},\qqtext{etc.}
\end{equation}

\begin{figure}
    \centering
    \includegraphics[width=\linewidth]{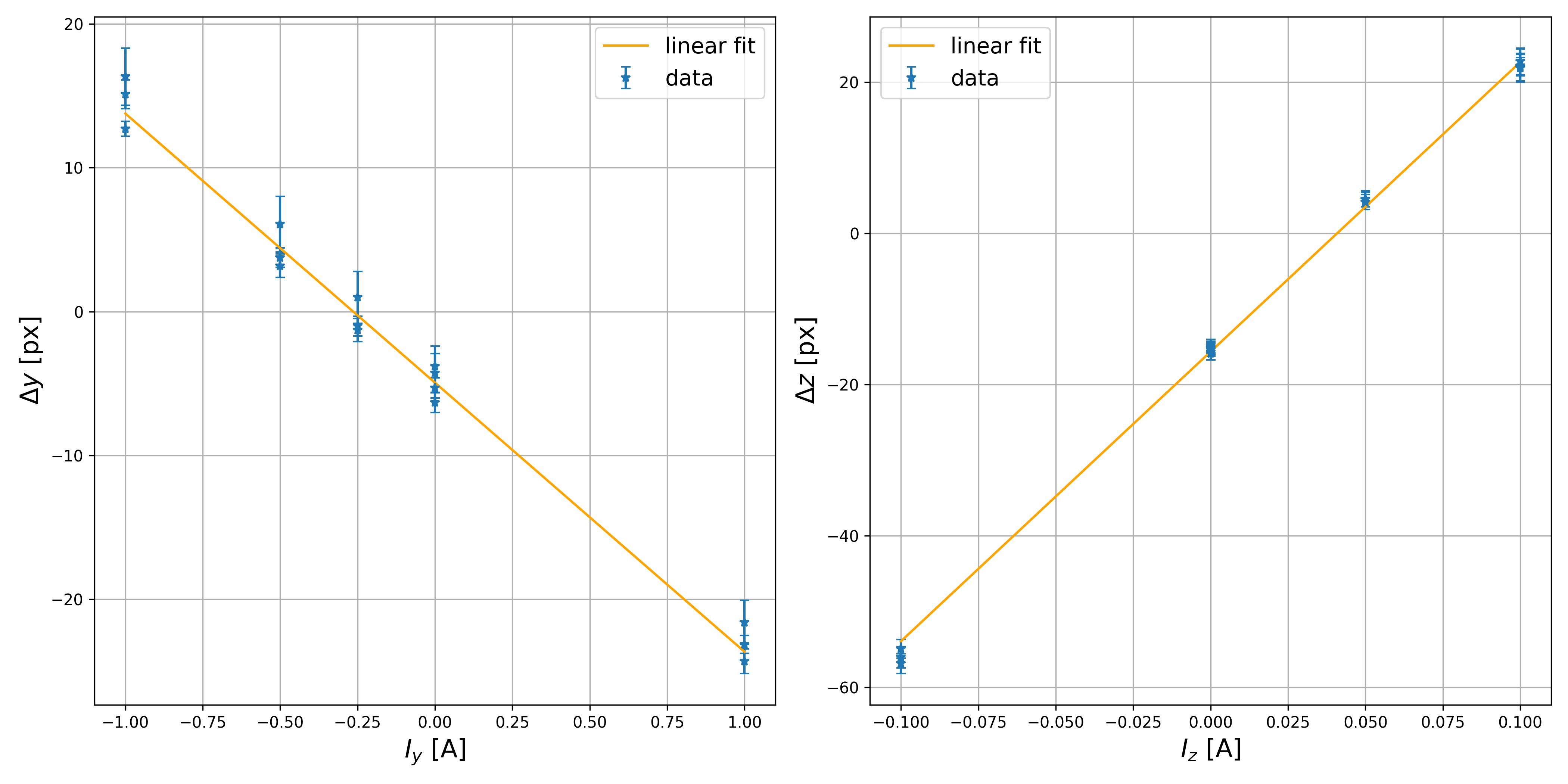}
    \caption{\textbf{Finding the correct compensation} Components of $\Delta \vb r$ as functions of the applied currents \( I_y \) and \( I_z \), extracted from the estimates of $\vb r_0^{(+)}$ and $\vb r_0^{(-)}$, i.e., the large triangle positions in \cref{fig:results1,fig:results2}. The correct compensation current $I_y$ ($I_z$) is the zero-crossing – the $\Delta y=0$ ($\Delta z=0$) point – of the fitted linear function (orange line in each panel).}
    \label{fig:aggregate}
\end{figure}

\Cref{fig:aggregate} illustrates these considerations. The data is aggregated from both datasets presented in \cref{fig:results1,fig:results2}. A linear model can be fitted, whose zero crossings indicate the correct compensation currents as $I_y^@=\SI{-0.264(16)}{\ampere}$ and $I_z^@=\SI{0.0408(9)}{\ampere}$. Note that the spread of the inferred origins (rhombuses in \cref{fig:results1,fig:results2}) does not affect the precision of the compensation inference so long as it is due to common-mode effects.

\begin{figure}
    \centering
    \includegraphics[width=0.5\linewidth]{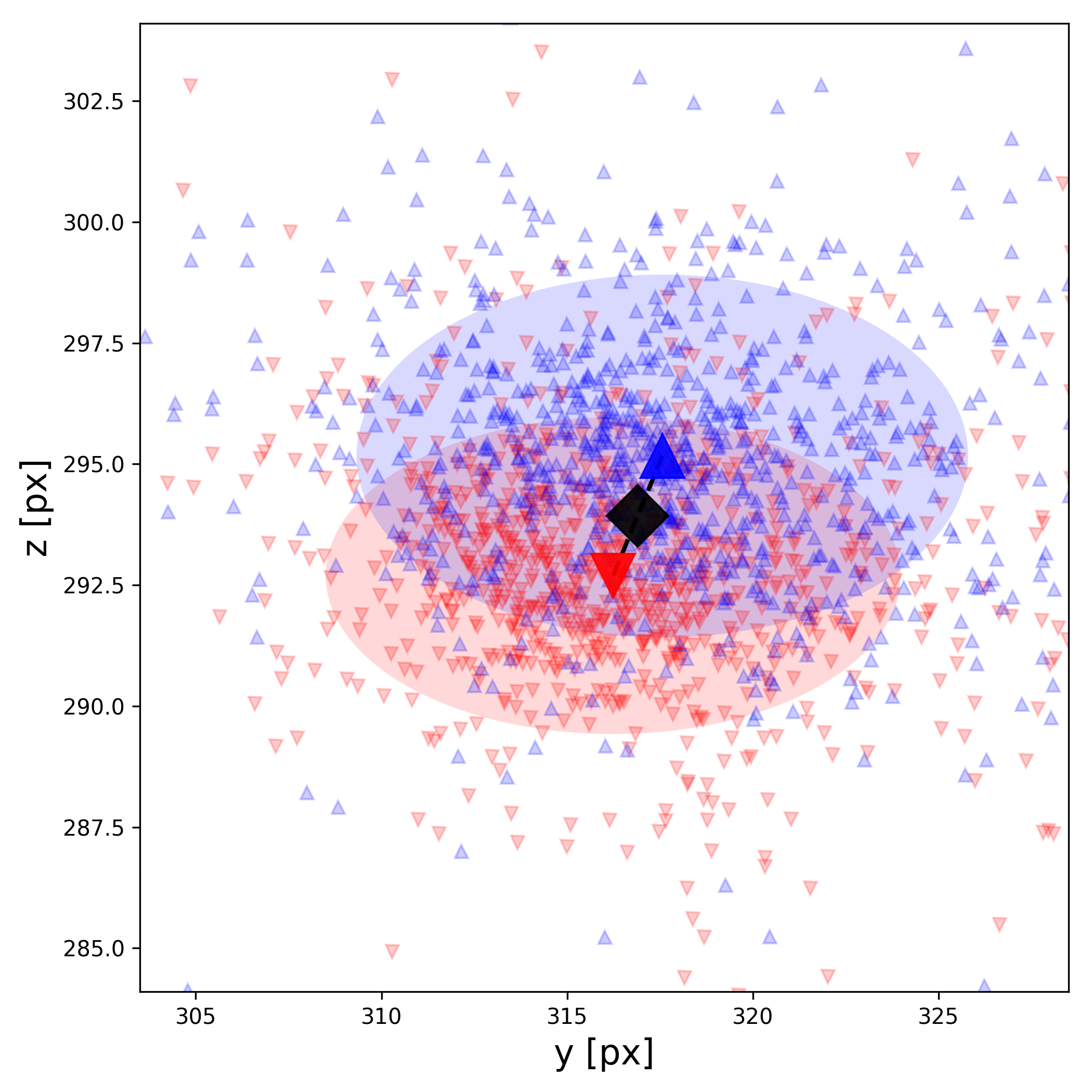}
    \caption{\textbf{The stopping of the buoy} at the compensation currents determined in \cref{fig:aggregate}, $I_y^@=\SI{-0.265}{\ampere}$ and $I_z^@=\SI{0.04}{\ampere}$. The figure shows the absorption imaging field of view with cloud centers stemming from individual experimental shots with alternating quadrupole polarities indicated with small triangles of opposite orientation as in \cref{fig:results1,fig:results2}, with the large triangles and the rhombus being the estimates of $\vb r_0^{(+)}$, $\vb r_0^{(-)}$, and $\vb 0$, respectively. The ellipses depict standard deviations of the positions in the ensembles for each polarity. Here, for the sake of visualizing the two largely overlapping ensembles, we used red and blue colors for plotting the opposite polarities.}
    \label{fig:stopping}
\end{figure}

The stopping of the buoy, that is, the independence of the trap center on quadrupole polarity in case of $\vb B_\text{ext}\approx0$ is further demonstrated in \cref{fig:stopping}.

\subsection{Precision}
\label{sec:precisionMetro}
To estimate the precision of the cold-atom buoy technique, we need to address how precisely the center of a magnetically trapped cold ensemble can be determined, since the technique is based on identifying this with the center of the quadrupole itself – which is possibly shifted by external fields. Any uncertainty in position can be directly translated to an uncertainty of magnetic field via the known quadrupole gradient, whose unit is magnetic field per unit length.

The uncertainty of the cloud center has statistical and systematic components. The former can be decreased by increasing the statistical ensemble that we use for averaging, while the second can manifest in different magnitudes on different time scales. The interplay of these errors can be demonstrated with the Allan deviation used in various cold-atom contexts \cite{krzyzanowska2023matter,zhu2023efficacy,zheng2024reducing}.

\begin{figure}
    \centering
    \includegraphics[width=0.8\linewidth]{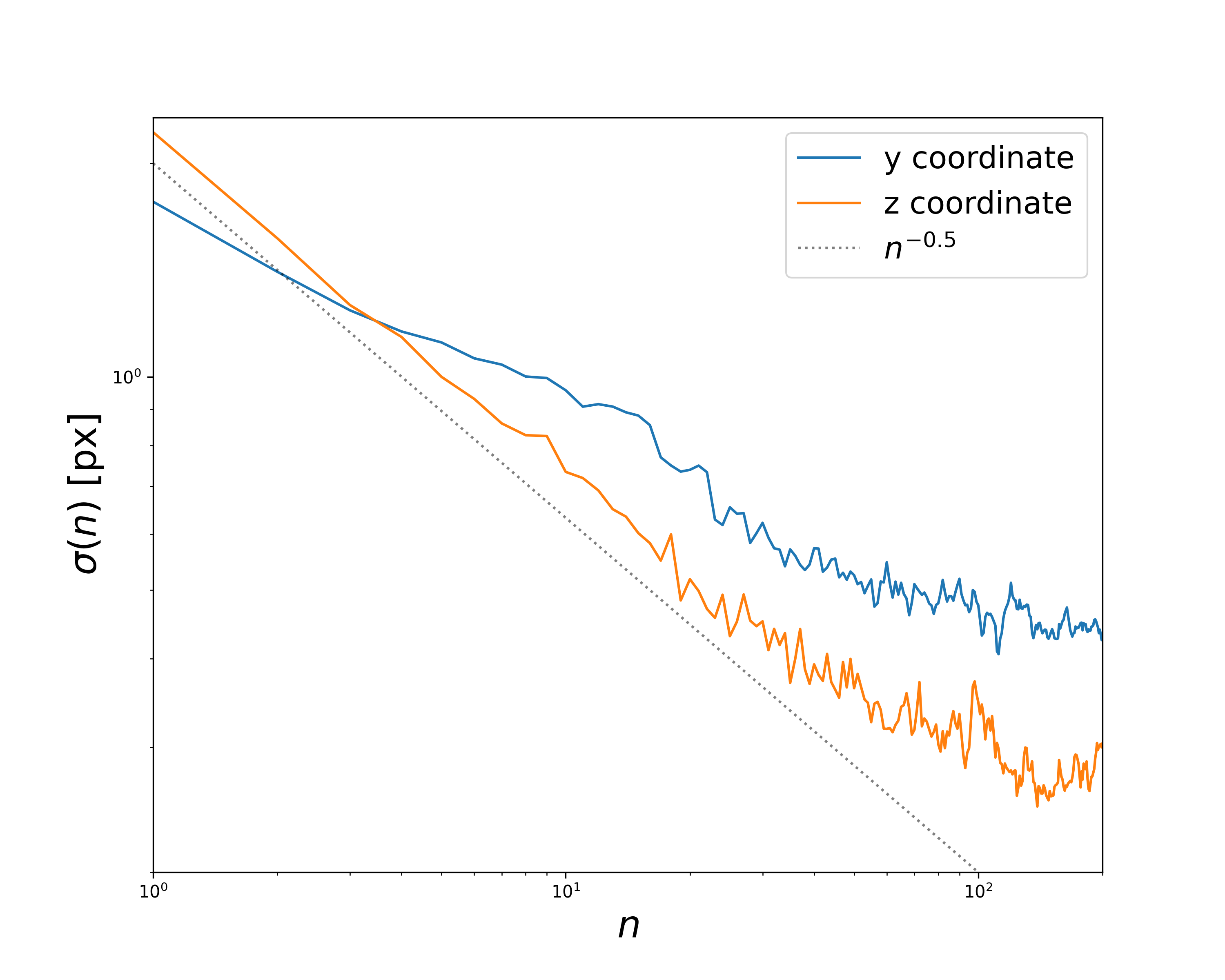}
    \caption{\textbf{Non-overlapping Allan deviation.} Statistical uncertainty of the determination of the cloud center as a function of ensemble size $n$, indicating the noise floor in our present system.}
    \label{fig:noiseFloor}
\end{figure}

An example is shown in \cref{fig:noiseFloor}, where data from a 12-hour-long measurement campaign is compiled. This ensemble of shots is partitioned into non-overlapping ensembles of size $n$ increasing from 1 to 200. The ensemble averages are calculated, and the standard deviation $\sigma(n)$ of these averages across ensembles is taken. In the log-log figure we indicated with dotted line the $n^{-0.5}$ dependence, whose slope corresponds to pure uncorrelated noise. While the uncertainty of the $z$ coordinate follows this initially, that of the $y$ coordinate has smaller initial slope, indicating systematic correlations in the noise, probably stemming from a periodic oscillation in the experimental system whose frequency becomes aliased due to the repetition rate of the experimental cycle.

Eventually both curves flatten into their respective noise floors. Since this is about twice as high in the $y$ direction ($0.4\;\text{px}$) than in the $z$ direction, whereas the $y$ gradient of the quadrupole field is half the $z$ gradient, the same uncertainty is obtained for the field – such an independence on the gradient is supported by a scaling argument below. Given the CCD pixel size of \SI{5}{\micro\meter}, the uncertainty of the $y$ coordinate of the center is $\sim\SI{2}{\micro\meter}$. The quadrupole gradient in these experiments were $\SI{2.5}{\Gauss\per\milli\meter}$, so the inferred field strength uncertainty reads
\begin{equation}
\Delta B = \frac{\partial B}{\partial y} \Delta y \sim \SI{5}{\milli\Gauss}.
\end{equation}

The inferred origin $\vb 0$ – rhombuses in \cref{fig:results1,fig:results2,fig:stopping} –, and the differential buoy signal $\Delta \vb r$ in \cref{fig:aggregate} both have statistical uncertainties reduced by a factor of $\sqrt{2}$ assuming uncorrelated noise in the estimation of $\vb r_0^{(+)}$ and $\vb r_0^{(-)}$. Nevertheless, the actual spread of rhombus positions observed in \cref{fig:results1,fig:results2} reflects not only the $\sigma(n)$ uncertainty of the cloud center, but also the systematics stemming from field inhomogeneities as discussed above, and additional systematics introduced by the variation of the compensation currents. In particular, settings far from optimal compensation degrade the MOT loading and influence the atom number and temperature, thereby degrading the determination of the cloud center in the magnetic trap.

Finally, let us consider the resolution of the technique as a function of the quadrupole gradient. The sensitivity, i.e., the $\Delta\vb r$ response to an external field scales simply as $Q^{-1}$, on the basis of \cref{eq:buoySignal}. This is very intuitive: the smaller the gradient, the larger the displacement of the $\vb B=0$ point under an external field. However, in a weaker trap, the cloud has a larger spread, the characteristic width scaling as: $\sigma\sim\frac{k_\text{B}T}{\mu Q}$, where $\mu$ is a characteristic magnetic moment. Taking into account the atom-number dependence of the absorption imaging \emph{contrast}, the smallest resolvable spatial displacement reads $\delta x\sim\frac\sigma{\sqrt N}$. Putting it together, we can see that the quadrupole gradient cancels from the field resolution: $\delta B\sim \frac{k_\text{B}T}{\mu \sqrt N}$, that hence depends on the temperature and atom number.

\section{Discussion}

\subsection{Common-mode rejection}
\label{sec:CMR}
In the discussion around \cref{eq:buoySignal} we have already seen that the uncertainty of the origin $\vb 0$ cancels as a common-mode effect in the differential signal $\Delta\vb r$.

There are two fundamental physical effects that affect the ideal buoy concept as depicted in \cref{fig:buoyScheme}, which are however also rejected as common modes. The first one is gravity that is added to the magnetic trap on the level of the potential. Hence, it does not alter the position of the minimum of the potential; rather, it renders the potential asymmetric: the gradient in the downward $z$ direction is smaller than in the upward. This results in a skewed cloud, making a 2D Gaussian underfitted. However, the effect of gravity is independent of the quadrupole polarity, therefore it does not affect the differential buoy signal $\Delta\vb r$. Note that the estimate of the origin $\vb 0$ is affected in the form of a shift in the downward direction. A possibility of a more refined model explicitly treating gravity in the estimates of $\vb r_0^{(+)}$ and $\vb r_0^{(-)}$ is discussed in \cref{sec:AI}.

The second such effect is the inhomogeneity of the external field. As we show in \cref{sec:inhomogeneities}, this affects the positions of the shifted centers as
\begin{equation}
\vb r_0^{(\pm)} = \mp \vb Q^{-1}\vb B_\text{ext}(\vb 0) + \vb Q^{-1}\vb G_\text{ext}\vb Q^{-1}\vb B_\text{ext}(\vb 0),
\end{equation}
to first order in the gradient matrix $\vb{G}_\text{ext}\equiv\left . \nabla \otimes \vb{B}_\text{ext}\right|_{\vb 0}$. The leading term is odd under polarity reversal and constitutes the ideal buoy response to the homogeneous part of the external field, while the first-order correction proportional to $\vb G_\text{ext}$ is even. Therefore, in the differential observable we find
\begin{equation}
\Delta \vb r = -\vb Q^{-1}\vb B_\text{ext}(\vb 0),
\end{equation}
i.e., to first order in the external-field inhomogeneity, the contribution from $\vb G_\text{ext}$ cancels in the buoy signal.

\begin{figure}
    \centering
    \includegraphics[width=1\linewidth]{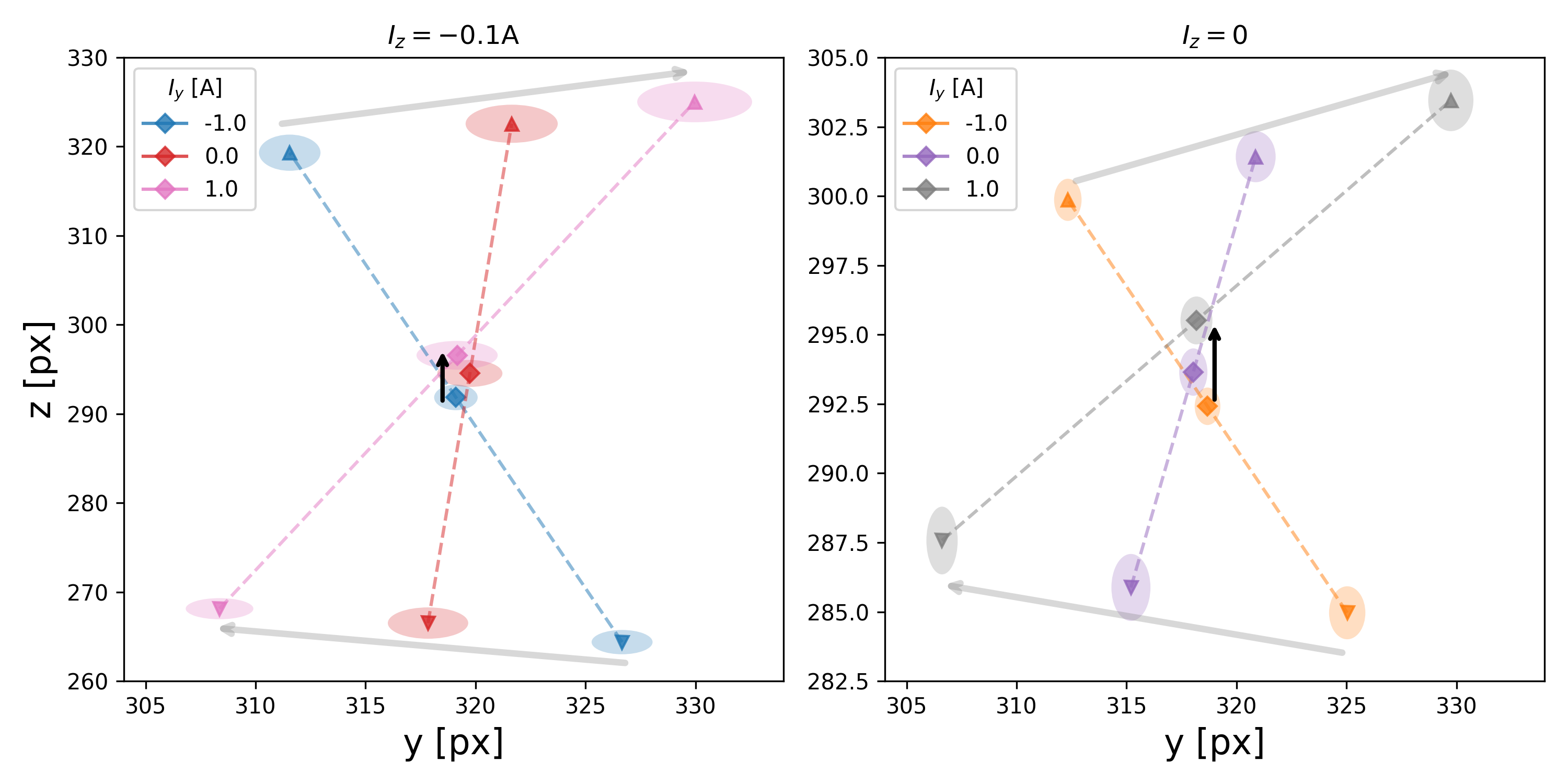}
    \caption{Subset of the buoy displacement data from \cref{fig:results1} for fixed $I_z=-0.1,\,\SI{0}{\ampere}$ and three values of $I_y$. The color code and notations are identical to that previous figure. The systematic drift of the inferred centers highlighted by the arrows indicates a common-mode displacement attributable to weak external magnetic-field inhomogeneity.}
    \label{fig:results1_small}
\end{figure}

In \cref{fig:results1_small}, we isolate a subset of the data shown in \cref{fig:results1}, corresponding to three values of $I_y$ while keeping $I_z=-0.1,\,\SI{0}{\ampere}$ fixed. In the ideal buoy picture, varying $I_y$ would induce a displacement solely along the $y$ direction. Instead, a residual displacement along $z$ is observed. The figure highlights that this additional displacement is a common-mode effect: the inferred center shifts monotonically upward as $I_y$ is varied. Moreover, this “walk of the buoy” is independent of the value of $I_z$, pointing towards the presence of a weak \emph{external} magnetic-field inhomogeneity that couples the spatial directions. In our system, the most probable source of this is the ion pump that is attached directly to the vacuum chamber.

\subsection{Extension to three dimensions via controlled inhomogeneity}
\begin{figure}
    \centering
    \includegraphics[width=0.5\linewidth]{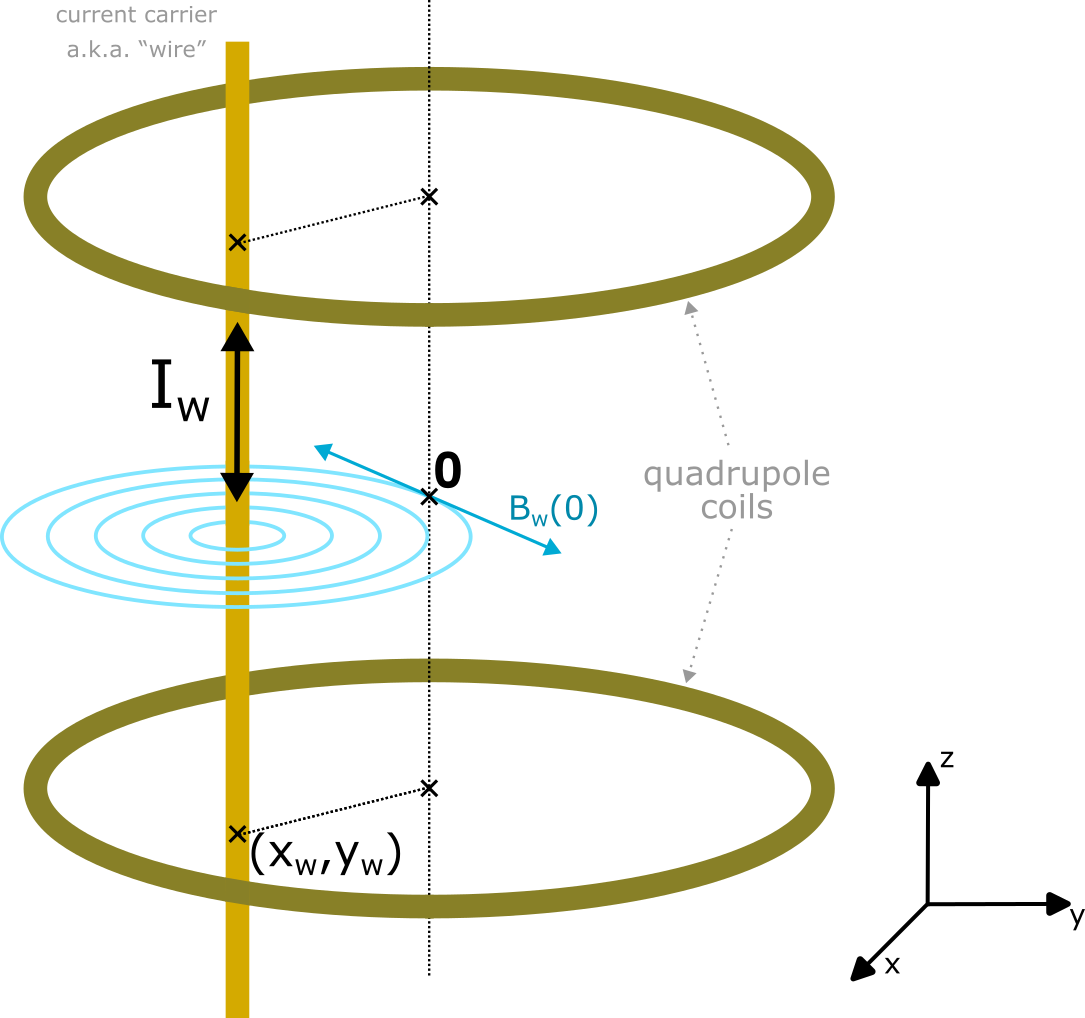}
    \caption{Schematic of the three-dimensional extension of the buoy technique using a controlled inhomogeneity. An additional current-carrying wire, aligned coaxially with the quadrupole, generates a magnetic field with both homogeneous and inhomogeneous components near the trap center. By reversing the current in the wire simultaneously with the quadrupole polarity, the homogeneous wire contribution becomes a common-mode effect and cancels in the differential buoy signal, while the induced off-diagonal gradients couple the response to the magnetic-field component along the imaging axis.}
    \label{fig:scheme3D}
\end{figure}

Let us now assume that the inhomogeneity is deliberately introduced by an additional electromagnet whose polarity can be controlled via the direction of its drive current. If the associated gradient matrix $\vb G$ contains suitable off-diagonal elements, such a controlled inhomogeneity can be used to couple the response of the buoy to the magnetic-field component along the imaging axis. The simplest realization of such a controlled inhomogeneity is an infinite current-carrying wire coaxial with the quadrupole, cf. \cref{fig:scheme3D}.

To illustrate the principle, we assume for simplicity that the external field is homogeneous. In the vicinity of the trap center, the total magnetic field can then be written as
\begin{equation}
\vb B(\vb r) = \vb Q\,\vb r + \vb B_\text{w}(\vb 0) + \vb G_\text{w}\,\vb r + \vb B_\text{ext}(\vb 0),
\end{equation}
where $\vb B_\text{w}(\vb 0)$ and $\vb G_\text{w}$ denote the homogeneous and inhomogeneous contributions of the wire field, respectively. In contrast to the uncontrolled inhomogeneity $\vb G_\text{ext}$ discussed in \cref{sec:CMR}, the wire gradient $\vb G_\text{w}$ is not required to be small; however, it is assumed to be controllable, for example by reversing the drive current of the electromagnet that generates it.

The resulting shift of the magnetic zero is
\begin{equation}
\label{eq:threeD}
\vb r_0 = \vb A\,\vb B_\text{w}(\vb 0) + \vb A\,\vb B_\text{ext}(\vb 0), \qqtext{with} \vb A \equiv -\big[\vb Q + \vb G_\text{w}\big]^{-1}.
\end{equation}
At first sight, the displacement appears to be contaminated by the homogeneous component of the wire field. However, consider the protocol in which the current through the wire is inverted simultaneously with the polarity reversal of the quadrupole. Under this combined operation, the full matrix $\vb A$ changes sign, while $\vb B_\text{w}(\vb 0)$ also changes sign. As a result, the contribution proportional to $\vb B_\text{w}(\vb 0)$ remains invariant and constitutes a common mode. Forming the differential buoy observable under this combined polarity reversal therefore simply yields
\begin{equation}
\Delta \vb r = \vb A\,\vb B_\text{ext}(\vb 0),
\end{equation}
where $\vb A$ can have the necessary off-diagonal terms to couple the field component  along the imaging axis to displacements in the imaging plane.

The above derivation is valid for a completely generic electromagnet configuration in addition to the quadrupole. For a straight wire oriented along the $z$ direction and located at a transverse position $(x_\text{w},y_\text{w})$ with respect to the quadrupole center, the relevant off-diagonal element of the gradient matrix reads
\begin{equation}
G_\text{w}^{(xy)}= -G_\text{w}^{(yx)} \propto x_\text{w}^2-y_\text{w}^2 ,
\end{equation}
so that the strength of the $x - y$ coupling depends only on the relative position of the wire in the transverse plane. For a fixed distance, this coupling is maximized when either $x_\text{w}=0$ or $y_\text{w}=0$, corresponding to placing the wire on one of the Cartesian axes.

In practice, choosing $y_\text{w}=0$ would make the wire intersect the imaging beam propagating along the $x$ direction, therefore the geometry $x_\text{w}=0$ is preferable. In this configuration, the controlled inhomogeneity generated by the wire maps the $x$ component of the external magnetic-field onto a displacement in the $y$ direction. This consideration motivates a two-step measurement protocol. First, the buoy technique is applied without the wire to compensate the homogeneous external field components in the $y$–$z$ plane. Subsequently, the wire is activated and its current is reversed simultaneously with the quadrupole polarity, enabling the remaining $x$ component of the external field to be inferred.

\subsection{Outlook}
The measurements presented in this work achieving sub-\SI{10}{\milli\Gauss} precision were performed without any specific optimization for metrological performance; they simply relied on our standard cold-atom stage and imaging configuration. We expect that another order of magnitude could be straightforwardly gained by targeted optimizations. Apart from some aspects of the imaging hardware, the statistical analysis can be improved. In this work, we employed a simple arithmetic mean to determine the average center position, i.e., the large triangles in \cref{fig:results1,fig:results2,fig:stopping}. While straightforward, this approach is sensitive to statistical outliers and may not represent the optimal center estimator under all conditions. To increase robustness against outliers \cite{thaprasop2021unsupervised}, alternative methods such as sigma clipping \cite{martin2025approach}, Chauvenet rejection \cite{maples2018robust}, or the Huber M-estimator \cite{huber2009} could be adopted.

In this connection we note that if the aim is restricted to \emph{sensing} external fields – specifically, detecting whether \emph{any} displacement occurs under quadrupole polarity reversal – the analysis can be simplified further. In such cases, unsupervised clustering methods such as \(k\)-means \cite{rodriguez2019identifying,zhang2025op} can be the choice to efficiently identify polarity-separated clusters. Conversely, if the aim is to find the correct compensation, we may not need large ensembles as in \cref{sec:demonstration}. Instead, we can iterate via Bayesian regression \cite{dose2003bayesian,von2011bayesian,martin2025approach}, feeding information directly into a linear fit similar to \cref{fig:aggregate}.

Calibration of the buoy technique to real units can be achieved analogously to the error estimation procedure in \cref{sec:precisionMetro}. As detailed in \cref{sec:precision}, we have a precise numerical model of the electromagnet configuration that allows for calculating the field in real units at any point within the experimental volume. By combining this with the known pixel size of the CCD we establish a direct correspondence in real units between field gradients and cloud displacements.

%second, via PGC efficiency / magnetic trap atom number.

\section{Methods}
\label{sec:Methods}

\subsection{System and protocol}
\label{sec:system}

\begin{figure}
    \centering
    \includegraphics[width=0.7\linewidth]{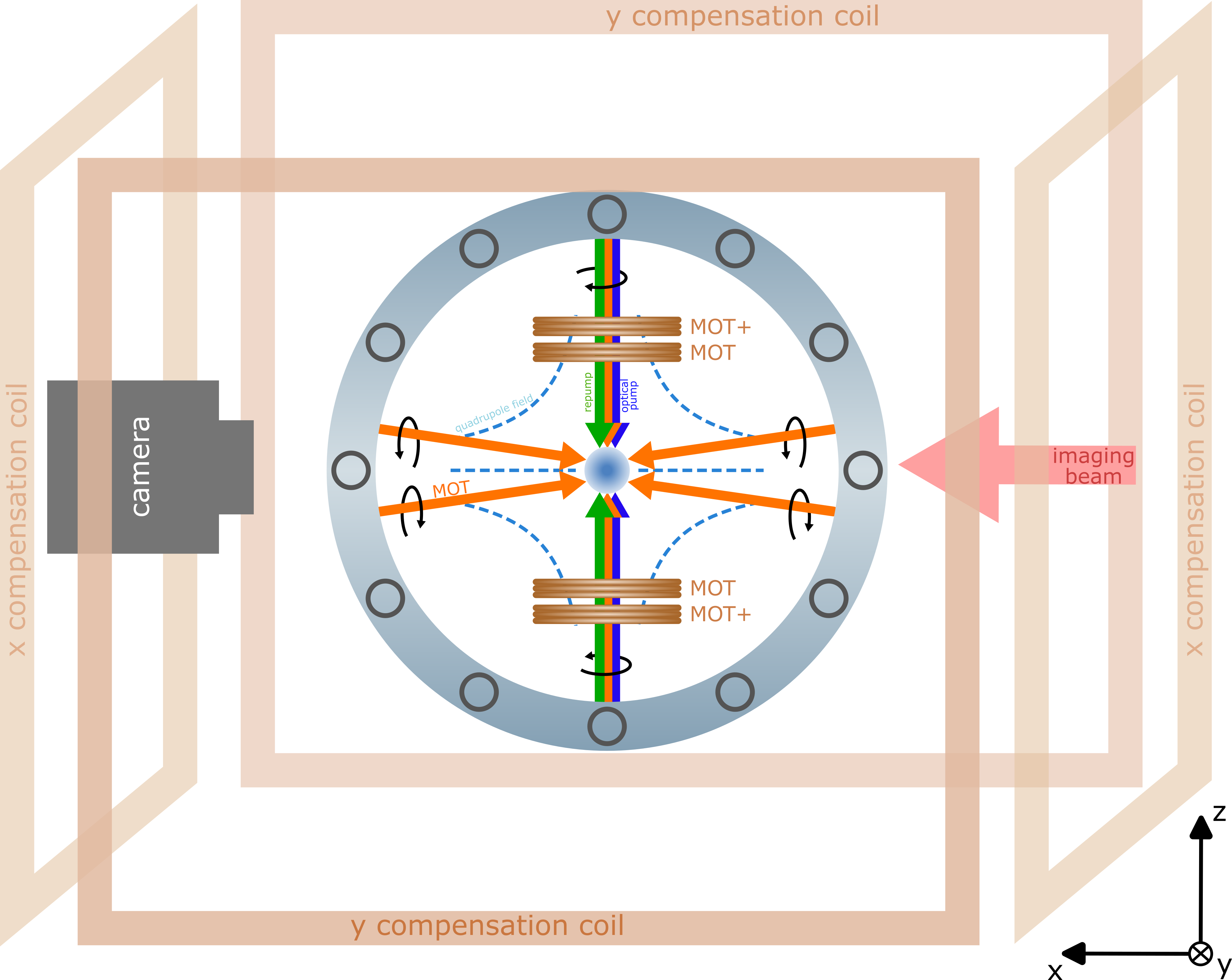}
    \caption{\textbf{The experimental setup.} The schematic viewport represents the vacuum chamber. The large rectangular loops around the chamber show the configuration of the external compensation coils along the \(x\) and \(y\) directions. Compensation in the \(z\) direction is provided by an additional intra-vacuo coil pair labeled MOT+.}
    \label{fig:setup}
\end{figure}

The experimental system is a rubidium-87 cold-atom system sketched in \cref{fig:setup}. The coils responsible for generating the MOT magnetic fields are mounted inside the vacuum chamber, allowing operation with relatively low currents – up to \SI{5}{\ampere} – for which ultra-precise current sources are available. The same in-vacuum MOT coil pair is also used for generating the magnetic quadrupole trap in a subsequent stage. The internal radius of the MOT coils is $\SI{32}{\milli\meter}$, their width $\SI{16}{\milli\meter}$, their thickness $\SI{10}{\milli\meter}$, and they are separated vertically by $\SI{34}{\milli\meter}$. They have 128 windings each in a stack of $16\times8$ (horizontal $\times$ vertical). The precision requirement for the current drivers is analyzed in \cref{fig:deviationFromHomogeneous}(b).

The experimental protocol proceeds as follows. After a MOT loading stage on the \(F=2 \leftrightarrow F'=3\) transition of the D$_2$ line, the magnetic quadrupole field is turned off, and polarization gradient cooling (PGC) is applied in the $\sigma^+ -\sigma^-$ configuration, cf. Sec. 8.4 of \cite{metcalf1999laser}, using the same optical beams as for the MOT. We then extinguish the MOT beams, retain only the repumper addressing the \(F=1\) ground state, and apply optical pumping on the \(F=2 \leftrightarrow F'=2\) transition in the presence of a small homogeneous magnetic field coaxial with the pump beam. It is to this field direction that the atomic magnetic moment becomes aligned during the pumping stage and remains so when the homogeneous field is adiabatically replaced by the quadrupole field at the onset of magnetic trapping. Ideally, the population is transferred into the low-field-seeking Zeeman sublevel \(F=2, m_F=2\) with respect to the quantization axis defined both by the beam axis and magnetic field.

Subsequently, a strong quadrupole magnetic trap is ramped up gradually to an axial gradient of \(5-\SI{6.6}{\Gauss\per\milli\meter}\). Here, except for a tiny radius around a magnetic zero traversing the cloud, the adiabaticity condition remains satisfied, i.e., the direction of the atomic magnetic moment follows the change of the local magnetic field. Ideally, the atoms remain in the low-field-seeking Zeeman sublevel \(F=2, m_F=2\) with respect to the \emph{local} magnetic field direction. After the trap reaches full strength, we wait for \(\SI{400}{\milli\second}\) to allow the atomic cloud to relax; that is, to damp center-of-mass and breathing oscillations before image acquisition.

The rubidium dispenser is located approximately \SI{1.5}{\centi\metre} from the MOT center. To minimize perturbations from its stray inhomogeneous magnetic field – arising from the current in its leads – it is operated in pulsed mode and switched off shortly before the end of the MOT collection stage. For full details of the setup and the experimental protocol cf. \cite{Varga2024}.

The magnetic trap, cf. Sec. 10.2 of \cite{metcalf1999laser}, provides an adiabatic potential since the atomic dipole has to follow the direction of the magnetic field at the position of the atom in order to remain in the low-field seeking state. Put otherwise, the parametric evolution (with parameter $\vb r$) of the Hamiltonian of the internal atomic dynamics has to remain adiabatic in order for the state to remain in $F=2, m_F=2$. The potential \labelcref{eq:potential} can be derived as:
\begin{equation}
\label{eq:potentialDerived}
    U(\vb{r}) = -\vb*{\mu} \cdot \vb{B}(\vb{r}) \approx - \qty( - \mu_\text{B} g_F m_F \frac{\vb{B}(\vb{r})}{\abs{\vb{B}(\vb{r})}} ) \cdot \vb{B}(\vb{r}) = \mu_\text{B} g_F m_F \abs{\vb{B}(\vb{r})},
\end{equation}
where the second, approximate equality expresses the adiabatic assumption, i.e., that $\vb*{\mu}$ is parallel to the local magnetic field $\vb B(\vb r)$; whereas the sign of $\vb*{\mu}$ reflects that of the low-field seeking state.

The three pairs of coils used to generate the compensation field \labelcref{eq:compensation} consist of (cf. \cref{fig:setup})
\begin{itemize}
    \item two large pairs dubbed compensation coils and mounted outside the vacuum chamber with their axes along the two horizontal directions ($x$ and $y$)
    \item an intra-vacuo pair dubbed MOT+ and mounted just outside of the MOT coil pair with its axis along the $z$ direction. They have the same geometry as the MOT coils, but are separated by $\SI{74}{\mm}$.
\end{itemize}
In \cref{eq:Bext}, $I_x$ and $I_y$ are the currents in the compensation coils, and $I_z$ is the one in the MOT+ coils. Since the MOT+ coil pair is much closer to the MOT center than the compensation coils, much smaller currents are needed for measurable displacement of the quadrupole trap, leading to the different scales in the two panels of \cref{fig:aggregate}.

\subsection{Absorption imaging and image processing}
\label{sec:AI}
Imaging is performed with a single lens outside the vacuum chamber, under the 2f condition. This yields the real size of the image, meaning that the pixel size in the absorption imaging field-of-view is the physical size of the CCD pixels, i.e. $\SI{5.3}{\micro\meter}$.

To determine the center-of-mass position of the atom cloud in the magnetic trap, we fit a two-dimensional Gaussian profile to the absorption image. In addition to the Gaussian core, the fit model includes a constant offset term to account for imperfections in the imaging process such as imperfect reference image and CCD dark current fluctuations. With these considerations, the fit is found to be robust, it converges reliably even with generic initial parameters.

To ensure that only reliable fits contribute to the statistical analysis, each image is subjected to a series of automated pre-registered quality-control checks. An image is retained only if all of the following conditions are met:
\begin{itemize}
%  \item reduced chi-squared of the fit $< 2$,
  \item uncertainty of the fitted cloud center $< 0.1\,\text{px} (\approx \SI{0.5}{\micro\meter})$,
  \item magnitude of residual skewness $< 1$,
  \item signal-to-noise ratio (Gaussian amplitude / constant offset) $> 10$.
\end{itemize}
In our experience, frames that fail these criteria occur in fewer than 1\% of cases and can be traced to timing errors arising from transient communication glitches between the CCD camera and the control PC.

It is worth noting that the quadrupole gradient can be used in an initial stage of magnetic trapping as a state-selective filter. By operating slightly above the threshold required to support atoms against gravity, only atoms in the maximally low-field-seeking Zeeman sublevel \(F=2,m_F=2\) remain confined, while atoms in \(m_F=1\) are lost. After such a filtering stage, the gradient may be increased to establish a stable magnetic trap for the buoy measurement. In that case, the ensemble occupies a single Zeeman sublevel, and its spatial distribution reflects thermal equilibrium in the corresponding tilted magnetic potential, where the point of maximal OD coincides with $\vb r_0$, the $\vb B=0$ point of the trap.

In the present work, however, no deliberate \(m_F\) filtering was applied. Due to imperfect optical pumping and possible nonadiabatic (Majorana) spin flips near the field zero, the trapped sample contains a mixture of Zeeman sublevels. Since the magnetic force is proportional to the magnetic moment, atoms in different \(m_F\) states experience slightly different effective trapping potentials and corresponding gravitational sag. The measured optical-density profile is therefore a superposition of these distributions. Empirically, this mixture yields a vertical density profile that is closer to a single Gaussian and can thus be more robustly fitted than the profile of a pure Zeeman-state ensemble. While a more refined fitting model including gravity and state-dependent sag could in principle be implemented, the differential observable \(\Delta \vb r\) remains unaffected by such common-mode effects.

\subsection{The effect of inhomogeneities}
\label{sec:inhomogeneities}
Identifying the spatial origin with the true center of the quadrupole, an inhomogeneous external field $\vb{B}_\text{ext}(\vb r)$ can be expanded in a Taylor series as
\begin{equation}
\vb{B}_\text{ext}(\vb r) = \vb{B}_\text{ext}(\vb 0) + \vb{G}_\text{ext} \vb r + \order{r^2},\qqtext{with}\vb{G}_\text{ext}\equiv\left . \nabla \otimes \vb{B}_\text{ext}\right|_{\vb 0}
\end{equation}
the external gradient matrix. Then, it is the sum of $\vb G_\text{ext}$ and the quadrupole matrix \labelcref{eq:Qpole} that determine the displacement of the magnetic zero point, so that instead of \cref{eq:displacedCenter}, we have:
\begin{equation}
\label{eq:inhomogeneity}
    \vb r_0 = - \big[\vb Q + \vb{G}_\text{ext}\big]^{-1}\,\vb{B}_\text{ext}(\vb 0).
\end{equation}

Even though the total field still grows linearly with distance from the center and retains a single magnetic zero, several prerequisites of the ideal buoy effect are lost already at this lowest order of inhomogeneity. Most importantly, \( \vb r_0 \) no longer simply flips sign upon reversing the sign of \( \vb Q \), and the clean directional response to \( \vb{B}_\text{ext}(0) \) is also degraded. This is because, in contrast to the diagonal \( \vb Q \), the external gradient matrix \(\vb{G}_\text{ext}\) is generally non-diagonal.

%=>linearity in external current also lost.

To underpin the considerations of \cref{sec:CMR}, let us perform a small-inhomogeneity expansion in the right-hand side of \cref{eq:inhomogeneity}, in the case when the inhomogeneity of the external field is much weaker than that of the quadrupole, i.e. $\norm{\vb G_\text{ext}}\ll\norm{\vb Q}$:
\begin{equation}
\big[\vb Q + \vb{G}_\text{ext}\big]^{-1}=\vb Q^{-1} - \vb Q^{-1}\vb{G}_\text{ext}\vb Q^{-1} + \order{\vb{G}_\text{ext}^2}.
\end{equation}
Substituting this expansion into \cref{eq:inhomogeneity}, we obtain for the displaced magnetic centers
\begin{equation}
\vb r_0^{(\pm)} = \mp \vb Q^{-1}\vb B_\text{ext}(0) + \vb Q^{-1}\vb G_\text{ext}\vb Q^{-1}\vb B_\text{ext}(\vb 0) + \order{\vb{G}_\text{ext}^2}.
\end{equation}

\subsection{Estimating the precision of simple approximations}
\label{sec:precision}
The importance of magnetic trapping in cold-atom technology has long motivated the development of good approximate formulas for the trapping fields \cite{bergeman1987magnetostatic}. However, we can be numerically exact with the \texttt{magnetic} Python package \cite{magnetic} developed by our group. This package facilitates the calculation of static magnetic fields arising from arbitrary  electromagnet configurations. We created a faithful representation of our electromagnet configuration within the package.

\begin{figure}
    \centering
    \includegraphics[width=\linewidth]{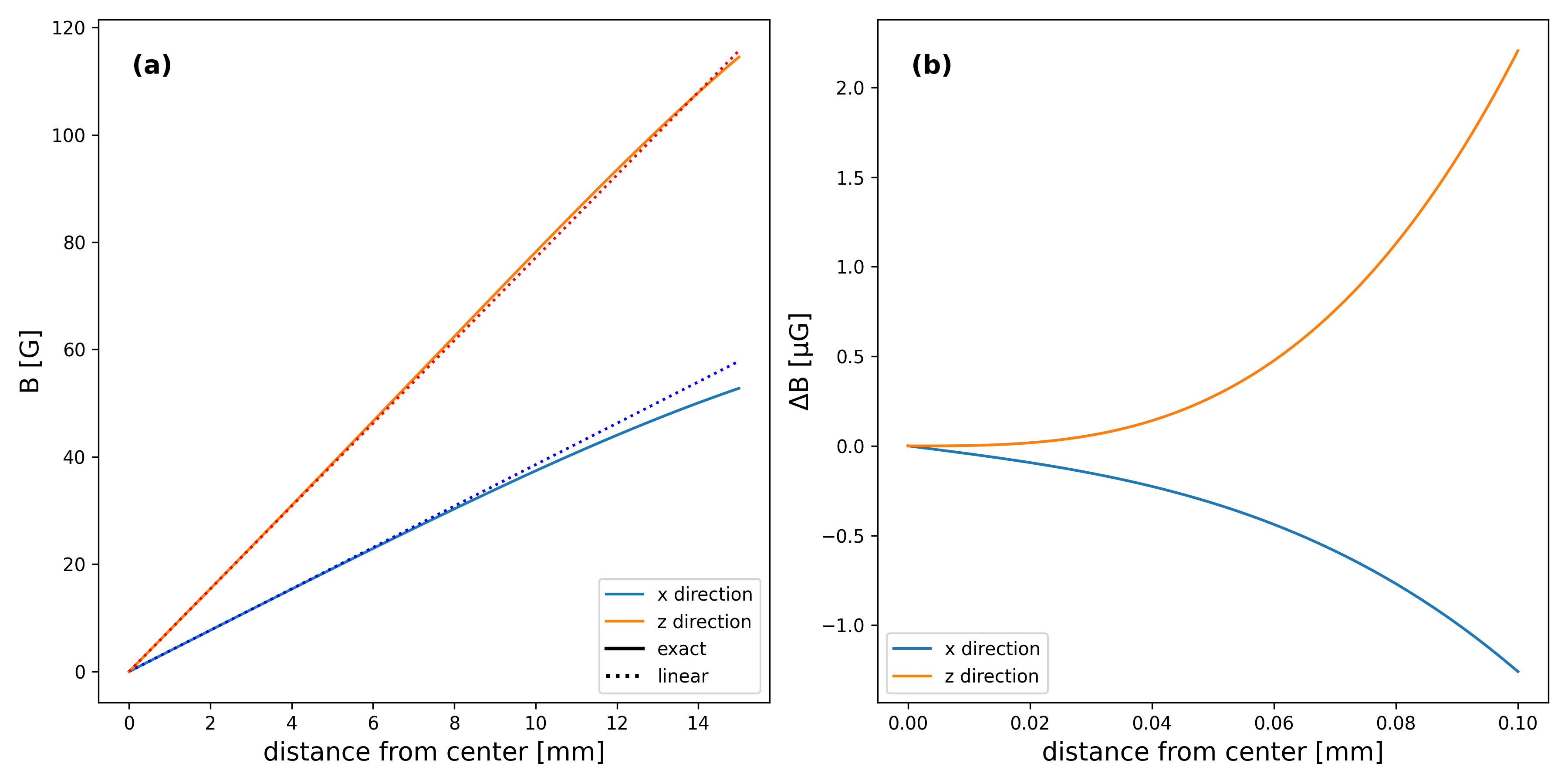}
    \caption{\textbf{Deviation of the magnetic field generated by the MOT coil pair from the ideal quadrupole form \labelcref{eq:Qpole}.} 
    (a) Comparison of the exact numerically computed field (solid lines) and the ideal quadrupole model (dotted lines), plotted along the horizontal (\(x\), blue) and vertical (\(z\), orange) axes. 
    (b) Difference between the exact and linearized fields. The deviation remains below the micro-Gauss level within a \SI{0.1}{\milli\metre} range around the trap center, supporting the validity of the quadrupole approximation at our current experimental resolution.}
    \label{fig:deviationFromQpole}
\end{figure}

As a benchmark, we investigate the validity of the ideal quadrupole \labelcref{eq:Qpole} in describing the actual field generated by the MOT coils. The comparison is shown in \cref{fig:deviationFromQpole}. On the largest spatial scale of \SI{0.1}{\milli\metre} ($\approx20\,\text{px}$) relevant to the displacement magnitudes in our experiment, the deviation from the ideal quadrupole shape is only at the micro-Gauss level. This is well below the resolution currently achievable with the buoy technique, confirming that the simple quadrupole model is sufficient for interpreting displacement data at our present precision.

\begin{figure}
    \centering
    \includegraphics[width=\linewidth]{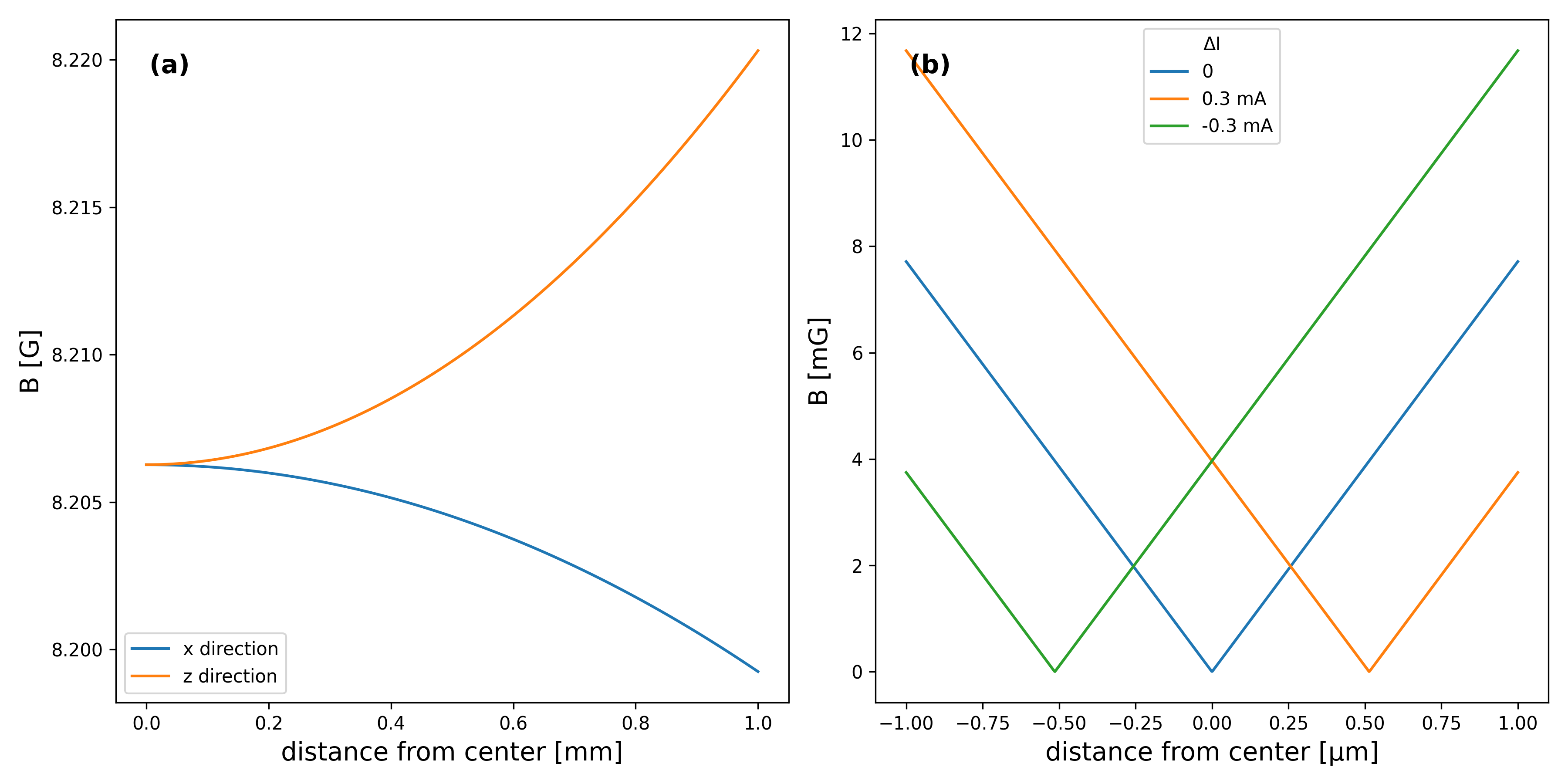}
    \caption{\textbf{(a)} Deviation of the magnetic field generated by the MOT+ coils from perfect homogeneity. The field variation reaches the milli-Gauss scale within $\SI{0.1}{\milli\metre}$ of the trap center along the axial ($z$) direction. \textbf{(b)} Simulated effect of common-mode fluctuations in the current source driving the MOT coil pair. A variation of $\pm\SI{0.3}{\milli\ampere}$ around the nominal \SI{4.7}{\ampere} results in an axial displacement of approximately $\SI{0.5}{\micro\metre}$ in the magnetic potential minimum, corresponding to a shift of about $0.1$ pixels.}
    \label{fig:deviationFromHomogeneous}
\end{figure}

\Cref{fig:deviationFromHomogeneous} illustrates two additional considerations relevant to the stability and accuracy of the buoy technique. Panel (a) quantifies the inhomogeneity of the bias field produced by the MOT+ coil pair along the $z$-axis. Even though the coils are operated symmetrically, the proximity and geometric constraints introduce inhomogeneity, resulting in variations of up to \SI{1}{\milli\Gauss} over a spatial extent of \SI{0.2}{\milli\metre}. This magnitude would impact the technique before reaching an order of magnitude improvement in resolution. The workaround would be to install a large external coil pair for the $z$ compensation as well. Panel (b) examines the susceptibility of the trap center position to common-mode current fluctuations in the MOT coils. Our simulations show that even a \SI{0.3}{\milli\ampere} deviation induces less than a micron-scale displacement. Since our current sources have a stability on the order of \SI{1}{\micro\ampere}, these results confirm that current noise contributes negligibly to the observed displacements.

\section{Acknowledgments}
This research was supported by the Hungarian National Research, Development and Innovation Office (Grant Nos. 2022-2.1.1-NL-2022-00004 and 2025-3.1.1-ED-2025-00011), the ERANET COFUND QuantERA programme (MOCA 2019-2.1.7-ERA\_NET-2022-00041), the QuantERA II Programme (V-mag 2024-1.2.2-ERA\_NET-2024-00012), and by the Swiss National Science Foundation (Grant No. 230870). AD and TWC acknowledges support from the János Bolyai research scholarship of the Hungarian Academy of Sciences.

\section{Data availability}
All absorption imaging data used in \cref{fig:results1,fig:results2,fig:aggregate,fig:stopping,fig:noiseFloor} will be made publicly available in the HUN-REN ARP Research Data Repository.

\bibliography{coldAtomBuoy}

\end{document}